\documentstyle[pra,epsf,aps]{revtex}

\def\db{\,\, {\bar{} \!\!d}\!\,\hspace{0,5pt}}
\def\cm#1{}
\def\lfrac#1#2{#1/#2}
\begin{document}
\sloppy
\title{Reparametrization Invariance of Path Integrals }
\author{H.~Kleinert\thanks{E-mail: kleinert@physik.fu-berlin.de} and
     A.~Chervyakov\thanks{E-mail: chervyak@physik.fu-berlin.de.
On leave from LCTA, JINR, Dubna, Russia. Supported by DFG under grant Kl 256.
                   }
	               ~\\Freie Universit\"at Berlin\\
          Institut f\"ur Theoretische Physik\\
          Arnimallee14, D-14195 Berlin}
\maketitle
\begin{abstract}
We demonstrate the reparametrization invariance
of
perturbatively defined
 one-dimensional
functional integrals up to
the three-loop level
for a path integral
of a quantum-mechanical point particle
in a box.
We exhibit the
origin of the failure of earlier authors
to establish
reparametrization  invariance
which led them to introduce, superfluously,
a compensating potential
depending on the connection of the
coordinate system.
We show that problems with
invariance are absent
by defining path integrals
as the
$ \epsilon \rightarrow 0$ -limit of
 $1+ \epsilon $-dimensional functional integrals.
\end{abstract}
~\\

{\bf 1}.~The long-standing problem of the reparametrization invariance
of
perturbatively defined
one-dimensional functional integrals, called path integrals,
has recently become
an important issue in the context of
worldline formulations of quantum
field theory \cite{1,2,3,3a}.
This problem
was apparently first encountered
in two-loop perturbative
calculations in Ref.~\cite{4}, and a
solution was attempted once more
in Ref.~\cite{5}. The latter authors exhibited
in great detail
the change of results
obtained after subjecting a path integral
 to a coordinate transformation.
The results could only be forced to
agree by
adding to the tranformed action
an artificial potential
term, of the order  $\hbar^2$,
depending on the connection of the coordinate transformation.

Such a noninvariance
under coordinate
transformations, if unavoidable,
would be extremely unpleasant
for two reasons:
First, the functional integral of the much more
complicated quantum gravity is known to be invariant
under coordinate transformations
if the infinities are regularized
a la t'{} Hooft and Veltman \cite{7} in $4- \epsilon $ dimensions,
and it would be surprising to see a failure
in simple quantum mechanics.
Second, a similar initially
encountered problem in the time-sliced definition of path integrals
has been solved elegantly in the textbook
(\ref{6}), by
defining it
as the image of
a euclidean path integral (in curved space via a nonholonomic
coordinate transformations).
 It would be embarrassing
if a perturbative definition
on a continuous time axis
which is standard in quantum field theory
 would fail to match the invariance of the time-sliced
definition.

An brief inspection of the problems encountered in
Refs.~\cite{4,5} reveals immediately the central unsatisfactory
feature of their treatment.
The perturbation expansion of the transformed path integral
leads to Feynman integrals involving
space-dependent kinetic terms
whose result depends on the order
of evaluation of the individual
one-dimensional momentum integrals. As a typical example,
take the Feynman integral
$$
 Y =
  \int  \frac{d k}{2\pi} \frac{d p_1}{2\pi}\frac{d p_2}{2\pi}
   \frac{k^2\,(p_1 p_2)}{(k^2 + m^2)(p_1^2 + m^2) (p_2^2 + m^2)
      [(k+p_1 + p_2)^2 + m^2]}.
$$
Integrating this  first over $k$, then over $p_1$ and
$p_2$ yields
$3/64m$. In the order first $p_1$, then $p_2$ and $k$, we find
$-1/64m$.
As we shall see below, the correct result is the average og the two, $1/32m$.

The purpose of this note is to exhibit the origin of this ambiguity
and to remove it by a proper definition
of the path integral as
the analytic continuation of a
corresponding functional integral in $D$ spacetime dimensions
to $D=1$. This leads
to a unique result exhibiting the desired
reparametrization invariance.
For the above example, the correct
result turns out to be the
 arithmetic-average of the two answers, i.e.,
$Y= 1/32m$ [to appear in the first term of
the decomposition of the
Feynman integral in Eq.~(\ref{@Eq39})].
The ambiguity will be removed by
a definite contraction
of the momentum vectors in the numerator.
A typical example characterizing the
resukting uniqueness
that the
integral
$$
 Y_0 =
  \int \frac{\db k}{(2\pi)^D}\frac{\db p_1}{(2\pi)^D}\frac{\db p_2}{(2\pi)^D}
   \frac{k^2\,(p_1 p_2)-(kp_1)(kp_2)}{(k^2 + m^2)(p_1^2 + m^2) (p_2^2 + m^2)
      [(k+p_1 + p_2)^2 + m^2]}.
$$
whose numerator vanished trivially in $D=1$ dimensions.
Because of the different contractions in $D$ dimensions, however,
$Y_0$ has a finite value
$Y_0=1/32m-(-1/32m)$
in the limit $D\rightarrow 1$, the result being split
according to the two terms in the numerator
[to appear in
the Feynman integrals (\ref{@Eq38}) and (\ref{@Eq39})].

\cm{This definition has
This permits us to ignore all diagrams which are divergent
like a delta function
$ \delta (0)=\int dp/(2\pi)$, and leads
to a unique result ghost diagrams and, at the same time, to find the correct results
for each of noncommutative diagrams in the limit of $D \rightarrow 1$.
These are always an arithmetic-average of different possibilities of the
one-dimensional noncommutative computation, that is, $G= 1/24$
 for the above diagram.}

To be specific, we consider a simple quantum-mechanical system,
 a point particle in a one-dimensional box of size $d$.
The ground-state energy of this system is exactly known,
$E^{(0)} =  \pi ^2 /2d^2$, and a method has been developed
to recover this value perturbatively \cite{8},
where a three-loop calculation yields the expansion
\begin{equation}
 E^{(0)} = \frac{m}{2} + \frac{g}{4} + \frac{1}{16} \frac{g^2}{m} +
    {\cal O}(g^3 ).
\label{0}\end{equation}
Here $g$ is a coupling constant which is equal to $ \pi ^2/d^2$, and $m$ an infrared regulator
to be taken to zero at the end.
This regulator comes about by replacing the confining box
by a smooth convex potential
which grows to infinity near the walls, and replacing the free-particle
euclidean action, the field energy, as follows:
\begin{equation}
E_0[u]=\int_{-L/2}^{L/2} dx\,  \dot u^2(x)\rightarrow
E[u]=\int_{-L/2}^{L/2} dx \left\{ \dot u^2(x)+m^2 g^{-1} \tan^2\left[ \sqrt{g} u(x)\right] \right\} ,
\label{@act}\end{equation}
where $L$ is the total euclidean
time interval. Being interested only in the ground state energy, this
may be assumed  to be very large, and  the limits of integration
will be omitted
from now on.
 For $m\rightarrow 0$, the initial box is recovered.
The $d\rightarrow \infty$,
the particle is completely free ($g=0$).
At finite $d$,
the potential is expanded in powers of $g$,
and the connected Feynman vacuum diagrams up to
three loops produce the perturbation expansion (\ref{0}).

The evaluation of the series (\ref{0}) requires a
conversion from the weak- to a strong-coupling expansion,
for which the techniques are now well developed \cite{6}
and described for this model in \cite{8}.

We shall now perform a
coordinate transformation
on this action and show that the ensuing
perturbation expansion leads
to the same expansion of the energy (\ref{0}).
As announced in the beginning,
a unique path integral will be obtained only
if it is defined as a $D\rightarrow 1$ -limit of a $D$-dimensional
system. The initial action will therefore be
extended accordingly, replacing $\int dx$ by $\int d^Dx$, and the kinetic term
$\dot u^2(x)$ by $\left[ \partial _\mu u(x)\right] ^2$.

{\bf 2}. The coordinate transformation
to be applied to
(\ref{@act}) is
\begin{equation}
u(x)\rightarrow \varphi(x)\equiv\sqrt{g} ^{-1}\tan\left[ \sqrt{g}u(x)\right]
.
\label{@transf}\end{equation}
The partition function of the transformed action in $D$ dimensions
is given by the euclidean path integral (in natural units)
\begin{equation}
  Z = \int  {\cal D} u (x)\,
e^{- \int d^Dx E_J[\varphi]}
e^{-\int d^Dx E[\varphi]},
\label{1}\end{equation}
 with the transformed field energy
\begin{equation}
    E [\varphi]= \frac{1}{2} \int d^Dx \left\{ [\partial_\mu u (\varphi (x))]^2 +
  m^2\, \varphi^2 (x) \right\} ,
\label{3}\end{equation}
and an effective field energy $ E_J[\varphi]$ coming from the
Jacobian of the coordinate transformation:
\begin{equation}
 E_J[\varphi]=
-\delta^{(D)} (0)\int d^Dx\frac{\partial u(x)}{\partial \varphi(x)},
\label{@jac}\end{equation}
where $ \delta ^{(D)}(0)\equiv \int d^Dp/(2\pi)^D$
is the inverse lattice volume associated with each  points in spacetime.

Inserting (\ref{@transf}) into (\ref{3}), and expanding
everything in powers of $g$,
the transformed energy becomes
\begin{equation}
E [\varphi]=E_m [\varphi]+E_{\rm int} [\varphi],
\label{5a}\end{equation}
with a harmonic part
\begin{equation}
E_m [\varphi]=
 \frac{1}{2} \int d^Dx \left\{ \left[ \partial _\mu\varphi(x)\right]^2+m^2 \varphi^2(x)\right\} ,
\label{5f}\end{equation}
and an interacting part
\begin{eqnarray}
E_{\rm int} [\varphi] &=& \frac{1}{2} \int d^Dx
 [\partial_\mu \varphi (x)]^2
 \left( - 2g \varphi ^2
     + 3 g^2  \varphi^4 - 4 g^3 \varphi^6 + \dots\right)
.\label{5}
\label{@}\end{eqnarray}

In the limit of large spacetime volume $L^D$,
the
harmonic part of the  partition function
gives an expression which can immediatly be taken to $D=1$ dimensions:
\begin{equation}
 Z_m = e^{- \rm Tr \log (\partial^2 + m^2)} =
   {\rm const.} \times e^{-L\lfrac{m}{2}},
\label{4}\end{equation}
 contributing the zeroth-order term $m/2$ to the expansion
(\ref{0}).
This is the exact result
for the partition function (\ref{1})
in the limit $d\rightarrow \infty$,
where the walls are absent.

For a finite distance $d$,
we perform a perturbation expansion
in powers of $g$ around this $Z_m$
using Feynman diagrams.
By discarding all disconnected diagrams,
the
expansion
applied directly to ground state energy.
 Since the interaction terms in (\ref{5})
have various forms,
there exists a variety of  Feynman diagrams. There are three types of
different
two-point functions,
\begin{equation}
 \Delta (x,x')=\langle \varphi (x) \varphi (x') \rangle = \int
    \frac{d^Dk}{(2 \pi)^D }
   \frac{e^{ik(x - x')}}{k^2 + m^2} =  \Delta (x,x'),
\label{6}\end{equation}
  and its derivatives
\begin{eqnarray}
 \Delta_\mu (x,x')&=&\langle \partial _\mu  \varphi (x) \varphi (x')\rangle
=
  \int \frac{d^Dk}{(2 \pi )^D}  \frac{ik^\mu}{ k^2 + m^2}
     e ^{ik (x - x')},\label{@vanish}\\
\Delta_{\mu \nu } (x,x')&=&\langle \partial _\mu   \varphi (x) \partial_ \nu  \varphi (x')\rangle  =
 \int \frac{d^Dk}{(2 \pi)^D }  \frac{k^\mu  k^ \nu }{k^2 + m^2}
   e ^{ik (x - x')},
\label{7}\end{eqnarray}
which are represented by
three types of lines,
solid, dashed, and mixed:
\begin{eqnarray}
&&\hspace{0mm}\raisebox{-1mm}{\mbox{\input 1.tps }} =\frac{1}{p^2+m^2},~~~
\hspace{0mm}\raisebox{-1mm}{\mbox{\input 2.tps }} =\frac{p^\mu p^ \nu }{p^2+m^2},~~~
\hspace{0mm}\raisebox{-1mm}{\mbox{\input 3.tps }} =i\frac{p^\mu}{p^2+m^2}
\label{@}\end{eqnarray}
The  infinitely many interaction vertices are pictures as follows:
\begin{eqnarray}
&&\hspace{0mm}\raisebox{.6mm}{\mbox{\input 4.tps }} =g,~~~~~~~~~~~~\,
\hspace{0mm}\raisebox{-3.3mm}{\mbox{\input 4b.tps }} =g^2,~~~~~~~~~~~\,
\hspace{0mm}\raisebox{-3.3mm}{\mbox{\input 4b.tps }} =g^3,~~~~~\dots~~~~~.
\\\nonumber
\label{@}\end{eqnarray}

When calculating Feynman  integrals in momentum space of
arbitrary dimensions $D$,
the quantity $ \delta ^{(D)}(0)=\int d^Dk/(2 \pi)^D $
in the Jacobian energy  (\ref{@jac})
vanishes and can be dropped from the exponent in
(\ref{1}).
Thus we find, up to the three-loop level,
the following graphical expansion for the
free energy density
\begin{eqnarray}
f&=&-\frac{1}{2}\hspace{0mm}\raisebox{-1mm}{\mbox{\input 5.tps }}
-g\hspace{0mm}\raisebox{-1mm}{\mbox{\input 6.tps }} +\frac{9}{2}\,g^2\hspace{0mm}\raisebox{-3.2mm}{\mbox{\input 7.tps }}
\nonumber \\&&
-\frac{g^2}{2!}\left[ 4\hspace{0mm}\raisebox{-1.2mm}{\mbox{\input 8.tps }}
+2\hspace{0mm}\raisebox{-1.2mm}{\mbox{\input 9.tps }}
+2\hspace{0mm}\raisebox{-1.2mm}{\mbox{\input 10.tps }}
+4\hspace{0mm}\raisebox{-2mm}{\mbox{\input 11.tps }}
+16\hspace{0mm}\raisebox{-1.95mm}{\mbox{\input 12.tps }}
+4\hspace{0mm}\raisebox{-1.9mm}{\mbox{\input 13.tps }} \right] .
\label{Fig 2}\end{eqnarray}
At the two-loop level, there exists only one diagram
representing
the single Wick contraction $\langle \varphi \varphi \rangle
 \langle
 \partial \varphi \partial \varphi\rangle$
of the local
expectation value
$\langle \varphi^2
 ( \partial \varphi)^2\rangle$.
At the three-loop level, the diagram $\!\!\!$\hspace{0mm}\raisebox{-1.7mm}{\mbox{\input 7s.tps }}
pictures one of the three equal nonzero
Wick contractions
$\langle \varphi \varphi \rangle  \langle \varphi \varphi \rangle
 \langle \partial \varphi \partial \varphi\rangle$
of the local
expectation value
$\langle \varphi^4 (\partial\varphi)^2\rangle$.
The remaining three-loop diagrams are either
of the three-bubble type, or of the watermelon type,
each with all possible combinations of
the three line types.
Since
the local expectation value $ \langle \partial \varphi \, \varphi\rangle$
vanishes according to Eq.~(\ref{@vanish}), several three-bubble diagrams
do not contribute. Only those which do are shown in Eq.~(\ref{Fig 2}).

{\bf 3}. All diagrams of Eq.~(\ref{Fig 2}) converge in momentum space,
 thus requiring no regularization to make them finite, as we would expect
for a quantum mechanical system.
Trouble arises, however, in some multiple
momentum integrals, since they
yield different results depending
on the order of their evaluation.
The ambiguity discussed in the beginning
occurs in the diagram is $\!\!\!$\hspace{0mm}\raisebox{-2.1mm}{\mbox{\input 12.tps }}$ $.
This and related diagrams require
a calculation in $D  $ dimensions, and going to
the limit  $D\rightarrow 1$.

The diagrams
which need a careful tratment are easily recognized in configuration space,
where the one-dimensional
propagator (\ref{6}) is
the continuous function.
 $ \Delta(x,x')=(1/2m) e^{-m|x-x'|}$.
Its first derivative has a jump at equal arguments
and is a distribution which is unproblematic
as long as the remaining integrand does not contain
$ \delta $-functions or their derivatives.
These appear with
second
derivatives
 of $ \Delta (x,x')$, where
the $D$-dimensional evaluation
msut be invoked to obtain a uniwue result.

The loop integrals encountered in $D$ dimensions
are based on the
basic one-loop integral
\begin{equation}
 I \equiv \int \frac{\db^D k\,}{k^2 + m^2}  = \frac{(m^2)^{\lfrac{D}{2}-1}}
        {(4 \pi )^{D/2}}   \Gamma \left(1 - \lfrac{D}{2}\right)
     \mathop{=}_{D\rightarrow 1}
			       \frac{1}{2m},
\label{8}\end{equation}
%
where we have set $\db ^D k\, \equiv d^D k /(2 \pi )^D$, for brevity.
For $D=1$, $I$ is equal to
$1/2m$.
By differentiation with respect to $m^2$ we can easily generalize
(\ref{8}) to
\begin{equation}
  I_ \alpha ^ \beta  \equiv  \int \frac{\db^D k (k^2)^ \beta }{(k^2 + m^2)^ \alpha }
     = \frac{(m^2)^{ \lfrac{D }{2}+ \beta - \alpha }}
{(4 \pi )^{D/2}}
      \frac{ \Gamma \left(\lfrac{D}{2} +  \beta \right)  \Gamma
         \left( \alpha - \beta -\lfrac{D}{2}\right)}{ \Gamma \left(\lfrac{D}{2}
      \right) \Gamma ( \alpha )} .
\label{9}\end{equation}
Note that $I_0^ \beta =\int \db^D(k^2)^ \beta =0$.
With the help of Eqs. (\ref{8}) and (\ref{9}) we calculate immediately
the local  expectation values (\ref{6}) and (\ref{7})
and thus the local diagrams in (\ref{Fig 2}):
\begin{equation}
\!\!\!\!\hspace{0mm}\raisebox{-1mm}{\mbox{\input 6.tps }} \!=
\langle \varphi^2\rangle\langle \partial \varphi\,\partial \varphi \rangle =
  \int \frac{\db^D q}{q ^2 \!+\! m^2}
     \int \frac{\db^D p \,p^2 }{ p^2 \!+\! m^2}   \mathop{=}_{D\rightarrow 1}
-
    \frac{1}{4},\!\!\!\!\!
\label{10}
  \!\!\!\!\!\! \hspace{6mm}\raisebox{-3.2mm}{\mbox{\input 7.tps }}\! = \langle \varphi\varphi\rangle ^2 \langle \partial
    \varphi \partial \varphi\rangle =
 \left( \int \frac{\db^D q}{q ^2 \!+\! m^2}\right)^2 \!\!
     \int \frac{\db^D p \,p^2 }{ p^2 \!+\! m^2}   \mathop{=}_{D\rightarrow 1}
 - \frac{1}{8m}.
\label{11}\end{equation}
  The three-bubble diagrams
in
Eq.~(\ref{Fig 2}) can also be easily computed.
If the
external bubbles are identical,
we obtain
\begin{eqnarray}
  \!\!\!\!\!\!\!\!\!\!\!\!\!
\hspace{0mm}\raisebox{-1mm}{\mbox{\input 9.tps }} & = & \int d^D1\, \Delta (1,1) \Delta _{\mu \nu }^2(1,2)
  \Delta (2,2)
= \left( \int \frac{\db^D q}{q ^2 \!+\! m^2}\right)^2
     \int \frac{\db^D p (p^2 ) ^2}{ (p^2 \!+\! m^2)^2}
\mathop{=}_{D\rightarrow 1}
    - \frac{3}{16m},
\label{12}\\
 \hspace{0mm}\raisebox{-1mm}{\mbox{\input 10.tps }} & = &\int d^D1 \, \Delta _{\mu \mu }(1,1)
 \Delta ^2 (1,2)
 \Delta _{\nu \nu }(2,2)
= \left( \int \frac{\db^D q \,q^2}{q^2 \!+\! m ^2} \right)^2
	  \int \frac{\db^D k }{(k^2 \!+\! m ^2)^2}
\mathop{=}_{D\rightarrow 1}
	\frac{1}{16m},
\label{13}\\
 \hspace{0mm}\raisebox{-1.5mm}{\mbox{\input 8.tps }}  &=&
\int d^D1 \, \Delta (1,1)
 \Delta_\mu ^2 (1,2)
 \Delta _{\nu \nu }(2,2)
	   =
 \int \frac{\db^D k}{k ^2 + m^2} \, \int
    \frac{\db^D p \, p^2}{ (p^2 + m ^2)^2}
 \int  \frac{\db^D q q ^2}{q ^2 + m ^2}
\mathop{=}_{D\rightarrow 1}
 - \frac{1}{16m}.
\label{14}\end{eqnarray}

We now turn to the  watermelon-like diagrams
which are more tedious to compute.
They require  a further  basic integral \cite{9}
\begin{eqnarray}
 J(p) & = & \int \frac{\db^D k\,}{ (k^2 + m ^2) [(k+p)^2 + m^2]}
=\int^{1}_{0} dx \int \frac{\db ^D k\,}{ \left[ k^2 + p ^2 x
     (1-x) + m^2 \right] ^2}
\nonumber \\  & = &
\frac{ \Gamma \left(2- \lfrac{D}{2}\right)}{(4 \pi )^{D/2}}
	\, \left(\frac{p^2 + 4 m^2}{4}\right)^{\lfrac{D}{2} - 2}
{}_2F_1 \left( 2 - \frac{D}{2}, \frac{1}{2};
     \frac{3}{2}; \frac{p^2}{p^2 + 4m^2}\right),
\label{15}\end{eqnarray}
where $
{}_2F_1
$ is the hypergeometric function.
For $D=1$, the right-hand side is simply
$1/[m(p^2+4m^2)]$. Differentiating $J(p)$
repeatedly with
respect to $p_\mu$, we obtain the
generalizations of $J(p)$:
\begin{equation}
  J^{\mu_1  \dots \mu_n} = \int \frac{\db ^D k\,~ k^{\mu_1} \cdots
        k^{\mu_n}} {(k^2 + m^2) [(k+p)^2 + m^2 ]}  ,
\label{16}\end{equation}
of which we need
the special
cases
\begin{equation}
  J^\mu (p) = \int \frac{\db^D k\, \, k^\mu}{(k^2 + m ^2)
             [(k + p )^2+m^2] [(k+p)^2 + m^2 ] }  = -\frac{1}{2}
    \, p ^\mu \, J (p),
\label{17}\end{equation}
and
\begin{eqnarray}
  J^{\mu \nu } (p)  =  \int \frac{\db ^D k\, \, k^\mu k^ \nu }
      {(k^2 + m ^2) [(k+p)^2 + m^2] }
&=&   \left[  \delta ^{\mu \nu }
	     + (D-2) \frac{p^\mu  p^ \nu }{p^2} \right]
    \frac{I}{2(D-1) } \label{18}\\
&+& \left[ -  \delta ^{\mu \nu } (p ^2 + 4m^2 ) +
    \frac{p^\mu p^ \nu }{p ^2} \left(D \, p^2 + 4 m^2\right)\right]
 \frac{J (p)}{ 4 (D-1)} ,\nonumber
\end{eqnarray}
whose trace is
\begin{equation}
 J_\mu{}^\mu (p) =  \int \frac{\db^D k\, \, k^2}{ (k^2 + m^2)
      [(k+p)^2 + m^2]} =I - m^2 \, J (p) .
\label{19}\end{equation}
We also encounter an integral
\begin{eqnarray}
  J_\mu{}^{\mu \nu } (p) &=& \int \frac{\db ^D k\, \, k^2 k^ \nu }
       { (k^2 + m^2)  [(k+p)^2 + m^2 ]} \nonumber \\&=&
\frac{1}{2} p^ \nu
	 [- I + m^2 J(p)].
\label{20}\end{eqnarray}
Various two- and three-loop integrals needed for the calculation
can be brought to the generic  form
\begin{equation}
   K (a,b) = \int \db ^D p \,\,(p^2)^a J^b (p) ,
	~~ a \geq 0,~~b \geq 1,~~ a \leq b,
\label{21}\end{equation}
and evaluated recursively as follows \cite{10}:
 From the Feynman parametrization of the first line of Eq.~(\ref{15})
 we observe that the two basic integrals (\ref{8}) and (\ref{15})
 satisfy the differential equation
\begin{equation}
  J(p) =  - \frac{\partial I}{\partial m^2} + \frac{1}{2}
     p^2 \frac{\partial J(p)}{\partial m^2} - 2p^2
     \frac{\partial J (p)}{\partial p^2}.
\label{22}\end{equation}
 Differentiating
 $K(a+1,b)$ from Eq.~(\ref{21}) with respect
to $m^2$,  and using Eq.~(\ref{22}), we find  the recursion
 relation
\begin{eqnarray}
K(a,b) =
\frac{2 b ({D}/{2} -1) \, I \, K (a-1,b-1) -2m^2
         (2a -2- b + D) K (a-1,b)}{(b+1) {D}/{2} - 2 b + a} ,
\label{23}\end{eqnarray}
which may be solved for increasing $a$ starting with
\begin{eqnarray}
 &&K(0,0) = 0,~~~~~
 K(0,1) = \int \db ^D p\, J(p) = I^2,\nonumber\\
  & & K(0,2) = \int \db ^D p\, J^2 (p) = A, ~\dots,~
\label{24}\end{eqnarray}
 where
 $A$ is the integral
\begin{eqnarray}
 A & =  & \int \frac{\db ^D p\,_1 \db^Dp_2 \db^D k\,}
         {(p^2_1 + m^2) (p^2_2 + m^2) (k^2 + m^2) [(p_1 + p_2 + k)^2
     + m^2] }
 =  \frac{1}{32 m^5}+{\cal O}(D-1).
\label{25}\end{eqnarray}
The one-dimensional limit
on the right-hand side follows directly from the
configuration space version of this integral
\begin{equation}
A=\int_{-\infty}^\infty dx\, \Delta ^4(x,0)
=\int_{-\infty}^\infty dx\, \left(\frac{1}{2m}e^{-m|x|}\right)^4=\frac{1}{32m^5}.
\label{@confA}\end{equation}

With the help of Eqs.~(\ref{23}) and (\ref{24}), we find
further integrals appearing in three-loop diagrams:
\begin{eqnarray}
 \int \db^D p\, p^2 J(p) &=&K(1,1)= - 2 m^2 I^2 ,
\label{26}      \\
 \int \db^D p\, \, p^2 \, J^2 (p)&=& K(1,2)  = \frac{4}{3}
	 (I^3 - m^2 A),
\label{27}\\
\int \db^D p\, (p^2)^2 J^2 (p) &=&  K(2,2) =  - 8m^2
       \frac{(6 -5D )I^3 + {2} Dm^2 A }{3(4-3D)}.
\label{28}\end{eqnarray}

We are now ready to calculate all the three-loop
contributions from the
watermelon-like diagrams in
Eq.~(\ref{Fig 2}). The first
without
mixed lines yields
\begin{eqnarray}
&&~~~\!\!\!\!\!\!\!\!\!\!\! \hspace{2mm}\raisebox{-2mm}{\mbox{\input 11.tps }}  =\int d^D1\,    \Delta ^2 (1,2)  \Delta _{\mu \nu }^2(1,2)
= \int \db^D p\, \db^D k\, \db ^D q
   \frac{(pk)^2}{(p^2 + m ^2)(k^2 + m ^2) (q^2 + m^2)
       [(p + k + q)^2 + m^2] } \nonumber \\
 & &~~~ ~~~\,~~~\!\!\!\!\!\!\!\mathop{=}_{q \rightarrow q \!-\! p}
 \int \db^D q \,J_{\mu \nu } (q) J^{\mu \nu } (q) =
    \int \db^D k\left. \frac{1}{16} (k^2)^2 J^2 (k)\right.\nonumber \\
&&\left.
\hspace{4.7cm}
+  \int \db^D k
 \frac{1}{4(D\!-\!1)}
      \left\{ D I^2 \!+\! \left[ (D-2) k^2 \!-\! 4m^2\right]
	I \, J (k) \!+\! \frac{1}{4} (k^2 \!+\! 4m^2)^2 J^2 (k)
    \right\} \right. \nonumber \\
  &&~~~ ~~~~~~\! =   - \frac{m^2}{2} \frac{(6 - {5} D) I^3
+ {2} D m^2 A}
          {3(4-3D)} - \frac{m^2}{6 (4-3D)}
	  \left[(6-5D) I^3
+2D m^2  A\right] \nonumber \\
&&~~~ ~~~~~~\!= - \frac{m^2}{3(4-3D)}
    \left[ (8-7D)I^3 + (D+4)m^2 A\right] \mathop{=}_{D\rightarrow 1}
  - \frac{m^2}{3} (I^3 + 5 m^2 A) = -\frac{3}{32m}.
\label{29}\end{eqnarray}
The diagram
with
two mixed lines
gives
\begin{eqnarray}
 & &~~\!\!\!\!\!\!\!\!\!\!\!\!\!~~~\!\!\!\!\!\!\!\!\!\!\!  \hspace{2mm}\raisebox{-1.9mm}{\mbox{\input 12.tps }} \!
 = \!-\!\!\int\! d^D1 \Delta (1,2)
 \Delta _\mu(1,2) \Delta _ \nu (1,2) \Delta _{\mu \nu }(1,2)
 \!= \!\!
 \int \!\db ^D k \db^D p_1 \db^D p_2 \frac{(kp_1)(kp_2)}
    {(k^2 \!+\! m^2) (p^2_1 \!+\! m^2) (p^2_2 \!+\! m^2) \left[
(k\!+\!p_1 \!+\! p_2)^2
     \!+\! m^2 \right] } \nonumber \\ &&
~~\!\!\!\!\!\!\!\!\!\!\!\!\!\!~~~ ~\,~~~\!\!\!\!\!\!\!\mathop{=}_{p_2 \rightarrow p_2 -k}
\int \db^D p\, \left[ p_ \nu  J_\mu(p) J^{\mu \nu } (p) + J_\mu
    (p) J_ \nu{} ^{\mu \nu }(p) \right] =
 - \frac{1}{8} \int \db^D p\, \, p^2 J(p) \left[
    (p^2 + 2m^2)J(p) - 2I\right] \nonumber \\
&&~~\!\!\!\!\!\!\!\!\!\!\!\!\!\!~~~ \,~~~~~\!= - \frac{m^2}{6(4-3D)}
\left[\left( 8-5D\right)I^3
       -  2 (4-D) m^2 A\right] \mathop{=}_{D\rightarrow 1}
 - \frac{m^2}{2} (I^3 - 2m^2 A) = -\frac{1}{32m}.
\label{@Eq38}\end{eqnarray}
Finally, the diagram with four mixed lines
is calculated by reduction
of the integral to the previous one:
\begin{eqnarray}
~~\!\!\!\!\!\!\!\!\!\! \hspace{2mm}\raisebox{-2mm}{\mbox{\input 13.tps }}  &=&
\int d^D1\Delta _\mu(1,2)
\Delta _\mu(1,2)
\Delta _\nu(1,2)
\Delta _\nu(1,2) \!
=-\!\!
 \int \db^D k \db^D p_1 \db^D p_2
   \frac{[k(k\!+\!p_1\!+\!p_2)]\,\,(p_1p_2)}
   { (k^2 \!+\! m^2)(p_1^2 \!+\! m^2)(p_2^2 \!+\! m^2) \left[ (k \!+\! p_1 \!+\! p_2)^2 \!+\! m^2\right]
   }
  \nonumber \\
 &=&-  \int \db^D k\, \db^D p_1 \db^D p_2
   \frac{k^2~(p_1 p_2)}{(k^2 \!+\! m^2)(p_1^2 \!+\! m^2) (p_2^2 \!+\! m^2)
      [(k\!+\!p_1 \!+\! p_2)^2 \!+\! m^2]}
\nonumber \\
&&- 2 \int \db^D k\, \db ^D p_1 \db^D p_2 \frac{(p_1 p_2)(kp_1)}
    {(k^2 + m ^2)(p_1^2 + m^2) (p_2^2 + m^2) \left[(k
       + p_1 + p_2)^2 + m^2 \right]  }\nonumber \\
&= &  - \frac{m^2}{3} (I^3 - m^2 A) + \frac{m^2}{3(4-3D)}
     \left[ (8-{5}D)I^3 -2 (4-D) m^2 A\right] \nonumber \\
 & &+ \frac{m^2}{3(4-3D)} \left[ (4-2D)I^3 - (4+D) m^2 A\right]
  \mathop{=}_{D\rightarrow 1}
 \frac{m^2}{3} (2 I^3 - 5 m^2 A) = \frac{1}{32m}.
\label{@Eq39}\end{eqnarray}
Note that since
the configuration space integral
of this diagram contains no $ \delta $-functions,
it could have been calculated immediately in one dimension
where $ \Delta_\mu (x,x')$ becomes  $-(1/2)\Theta(x-x')e^{-m|x|}$,
so that the integral over $d^D1$ reduces to $\int_{-\infty}^\infty dx [(1/2)e^{-m|x|}]^4=1/32m$.
~\\

{\bf 4}. Summing up all contributions of the various diagrams,
we find for the free energy density up to three
loops the perturbation expansion
\begin{eqnarray}
 f& = & \frac{m}{2} + \left(\frac{g}{4} - \frac{9}{16} \frac{g^2}{m}
   \right)- \frac{1}{2!} \left(-\frac{10}{8}\right)\frac{g^2}{m}
= \frac{m}{2} + \frac{g}{4} + \frac{1}{16} \,
      \frac{g^2}{m}.
\label{31}\end{eqnarray}
This is identical with the previous result (\ref{0}),
thus confirming the invariance of the perturbatively defined
path integral under coordinate transformations,
so that there is no
need for an artificial
potential term
of the order $\hbar^2$ called for by previous authors.

~\\
Acknowledgement:\\[1mm]
We are grateful to  Dr.~A.~Pelster for
useful contributions in the initial stages of this investigation,
and to M. Bachmann for useful discussions.

\end{document}